\documentclass[
reprint,
superscriptaddress,
 amsmath,amssymb,
 aps,
prb,
]{revtex4-1}

\usepackage{graphicx}% Include figure files
\usepackage{dcolumn}% Align table columns on decimal point
\usepackage{bm}% bold math
\usepackage{hyperref}% add hypertext capabilities
\usepackage{sidecap}
\usepackage{float}
\begin{document}
\title{Non-Abelian statistics in light scattering processes across interacting Haldane chains}
%\title{Evidence for non-Abelian statistics in light-induced triplet exchange within interacting Haldane chains}

% competiton proximity to a quantum critical point -> thermal versus quantum fluctuations show up as an unconventional thermal evolution fo multiparticle states .. predominantly observed in Raman scattering

\author{Vladimir Gnezdilov}
\affiliation{B. Verkin Inst. for Low Temperature Physics and Engineering, NASU, 61103 Kharkiv, Ukraine}
\affiliation{Inst. for Condensed Matter Physics, TU Braunschweig, D-38106 Braunschweig, Germany}

\author{Vladimir Kurnosov}
\affiliation{B. Verkin Inst. for Low Temperature Physics and Engineering, NASU, 61103 Kharkiv, Ukraine}

\author{Yurii Pashkevich}
\affiliation{O.O. Galkin Donetsk Inst. for Physics and Engineering, NASU, Kyiv - Kharkiv 03028, Ukraine}

\author{Anup Kumar Bera}
\affiliation{Helmholtz-Zentrum Berlin f\"ur Materialien und Energie, 14109 Berlin, Germany}
\affiliation{Solid State Physics Division, Bhabha Atomic Research Centre, Mumbai 400085, India}

\author{A.T.M. Nazmul Islam}
\affiliation{Helmholtz-Zentrum Berlin f\"ur Materialien und Energie, 14109 Berlin, Germany}

\author{Bella Lake}
\affiliation{Helmholtz-Zentrum Berlin f\"ur Materialien und Energie, 14109 Berlin, Germany}

\author{Bodo Lobbenmeier}
\affiliation{Inst. for Condensed Matter Physics, TU Braunschweig, D-38106 Braunschweig, Germany}

\author{Dirk Wulferding}
\affiliation{CCES, Inst. for Basic Science, Dept. Physics and Astronomy, Seoul Nat. Univ., Seoul 08826, Republic of Korea}
\affiliation{Inst. for Condensed Matter Physics, TU Braunschweig, D-38106 Braunschweig, Germany}
\affiliation{Lab. of Emerging Nanometrology LENA, D-38106 Braunschweig, Germany}

\author{Peter Lemmens}
\affiliation{Inst. for Condensed Matter Physics, TU Braunschweig, D-38106 Braunschweig, Germany}
\affiliation{Lab. of Emerging Nanometrology LENA, D-38106 Braunschweig, Germany}

\date{\today}

\begin{abstract}

The $S=1$ Haldane state is constructed from a product of local singlet dimers in the bulk and topological states at the edges of a chain. It is a fundamental representative of topological quantum matter. Its well-known representative, the quasi-one-dimensional SrNi$_2$V$_2$O$_8$ shows both conventional as well as unconventional magnetic Raman scattering. The former is observed as one- and two-triplet excitations with small linewidths and energies corresponding to the Haldane gap $\Delta_H$ and the exchange coupling $J_c$ along the chain, respectively. Well-defined magnetic quasiparticles are assumed to be stabilized by interchain interactions and uniaxial single-ion anisotropy. Unconventional scattering exists as broad continua of scattering with an intensity $I(T)$ that shows a mixed bosonic / fermionic statistic. Such a mixed statistic has also been observed in Kitaev spin liquids and could point to a non-Abelian symmetry. As the ground state in the bulk of SrNi$_2$V$_2$O$_8$ is topologically trivial, we suggest its fractionalization to be due to light-induced interchain exchange processes. These processes are supposed to be enhanced due to a proximity to an Ising ordered state with a quantum critical point. A comparison with SrCo$_2$V$_2$O$_8$, the $S=1/2$ analogue to our title compound, supports these statements.

%Physh keywords: Haldane, Floquet systems, Non-Abelian models, Transition-metal oxides, 1-dimensional systems, Raman spectroscopy

% three figures, 1 Table

%\begin{description}

%\item[PACS numbers]{75.50.Ee, 75.50.Mm, 75.40.Gb, 78.30.-y}
%\pacs{75.50.Ee, 75.50.Mm, 75.40.Gb, 78.30.-y}

%\end{description}
\end{abstract}

\maketitle

%%%%%%%%%%%%%%%%%%%%%%%%%%%%%%%%%%%%%%%%%%%%%
%\section{Introduction}
%%%%%%%%%%%%%%%%%%%%%%%%%%%%%%%%%%%%%%%%%%%%%

Topological effects in condensed matter physics are an attractive field of study as they combine very interesting concepts of fundamental physics with exotic effects, such as anomalous transport in magnetic fields, nonlocal interactions, and fractionalization of excitations ~\cite{haldane}. Prominent examples include relativistic Dirac electrons, quantum spin liquids, and the fractional quantum Hall effect.

One of the fundamental aspects of topological systems is that the exchange or braiding of quasiparticles in certain classes of systems leads to an unconventional phase, different from 0 or $\pi$, known from bosons and fermions. The resulting non-Abelian statistics has first been realized for the fractional Hall effect at 5/2 filling~\cite{wen91,moore91}. A planar system with a dimensionality D$<$3 has to be chosen. Furthermore, weakly interacting quasiparticles should exist that are degenerate in energy for large distances. This leads to the emergent quality of non-Abelian systems.

Periodically driving a system by an electric field is another approach to generate topological states~\cite{Khemani16,Moritomo16,Wang13,Esin18}. The so called Floquet Hamiltonian leads to energetically shifted replicas of electron-photon states. Here, the question appears whether the response to a Laser field may simultaneously induce a non-Abelian statistics by the exchange of quasiparticles~\cite{joel09,bishara09,Vogl-20}. Recently, it has been shown that such driving may enhance correlations~\cite{Peronaci-20,ESQFT} and transfer chain systems from Luttinger liquids to charge density wave instabilities~\cite{Kennes-18}. In principle, light-induced exchange processes are well known from insulating spin systems and described by the Loudon-Fleury Hamiltonian~\cite{Loudon68}. However, to our knowledge, the topological implications of light-induced spin exchange simultaneously with the Raman scattering process have not been addressed so far.

\begin{figure*}
	\label{figure1}
	\centering
	\includegraphics[width=12cm]{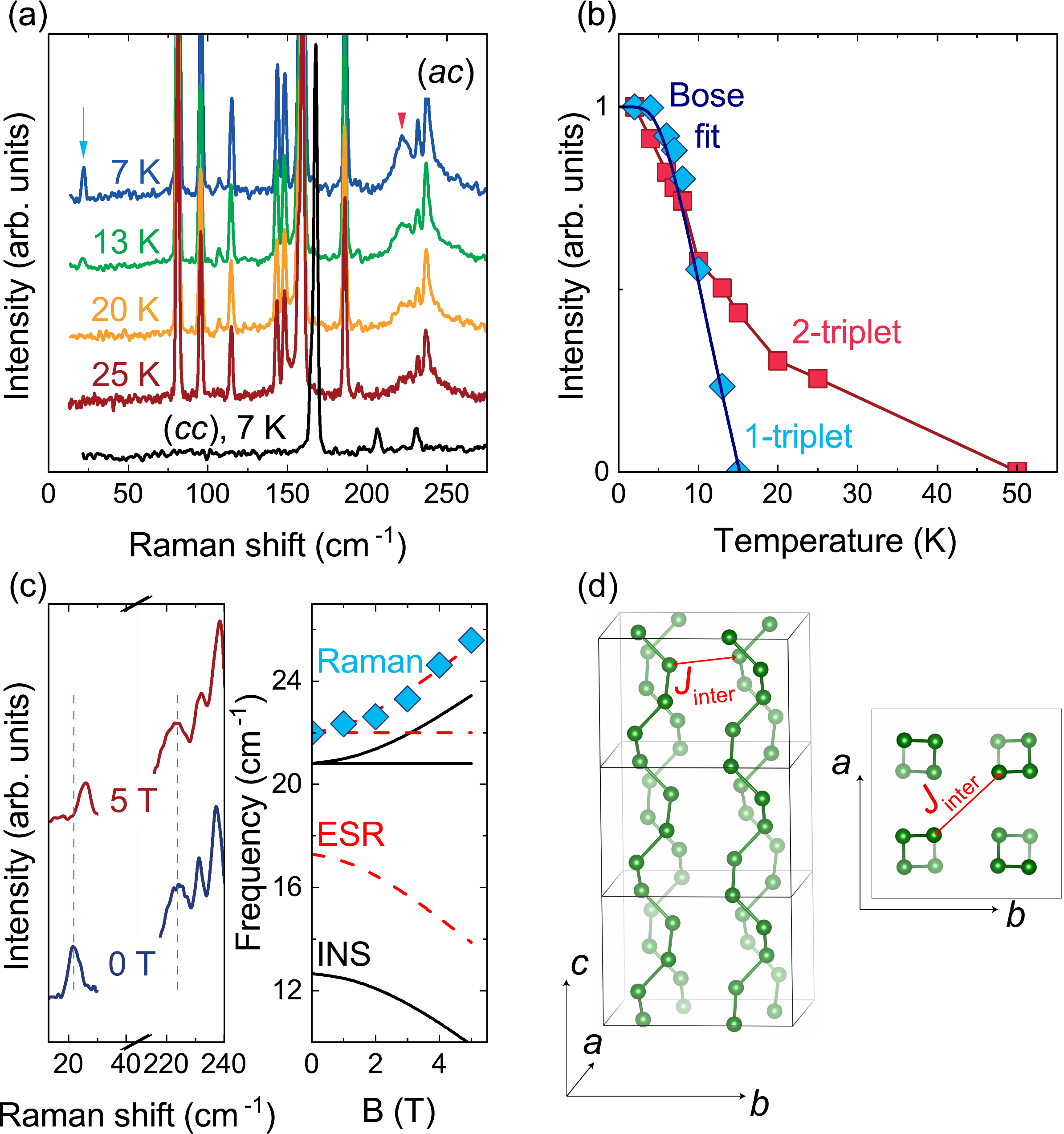}
	\caption{(Color online) (a) Temperature dependent Raman spectra of SrNi$_2$V$_2$O$_8$ obtained in ($ac$) polarization. The blue and red arrows mark the one- and two-triplet modes, respectively. For comparison, the 7-K spectrum in ($cc$) polarization is plotted in black. (b) Integrated intensity ($I(T)$) of the magnetic scattering contributions as a function of temperature. The blue line corresponds to a fit using a Bose factor. (c) Effect of a transverse magnetic field on the assigned one- and two-triplet modes. Zero field positions are marked by dashed lines. Transverse magnetic field effect on inelastic neutron scattering (full line), ESR (dashed line)~\cite{bera-13, bera-15, bera-15b, wang-13} and our one-triplet Raman mode (diamonds). (d) Four twisted Ni chains along the $c$ axis in two different projections. 	}
\end{figure*}

In the following report we focus on SrNi$_2$V$_2$O$_8$, a rather well-known representative of weakly interacting Haldane spin chains with $S$=1 states. This system has a singlet ground state with a spin gap in the bulk and topological states at the surface ~\cite{haldane}. No resonance effects exist in Raman scattering. Therefore, selection rules of conventional Raman scattering should be valid and follow the dominance of 1D correlations along the spin chains, i.e. appreciable intensity should only be observed with a polarization parallel to the chain direction (c axis). Nevertheless, finite interchain interactions and single ion anisotropies lead to Ising fluctuations and magnetic field induced long range order. With our work we follow the concept of inducing a topological state of an interacting chain system using the Raman scattering process and probe for possible non-Abelian statistics by the measurement process itself. A deconfinement of spinon and holon excitations with non-Abelian statistics has been proposed to exist for s=1 antiferromagnetic systems~\cite{greiner09}.

In SrNi$_2$V$_2$O$_8$ skewed spin chains are formed by edge-sharing NiO$_6$ (Ni$^{2+}$, $S = 1$) octahedra along the $c$ axis, see Fig. 1 (d). They are coordinated by VO$_4^{3-}$ (V$^{5+}$, $S = 0$) tetrahedra that weakly hybridize with the Ni-O coordinations. There are four chains per unit cell~\cite{bera-12}.

All important model parameters are well-known ~\cite{bera-13, bera-15, bera-15b, wang-13, bera-12, pahari-06, he-08} and the magnetic properties can be summarized as follows: (i) a Haldane gap of $\Delta=3.6$ meV (29 cm$^{-1}$) at the AFM zone center ($k = q_{\mathrm{chain}} = \pi$) separates the singlet, spin-liquid ground state from excited triplet states. The relation of the gap and the exchange coupling along the chain gives $\Delta_{0}$ = 0.41J = 0.41 $\cdot$ 8.7~meV = 3.57~meV (28.6~cm$^{-1}$).

From ESR two gaps are derived, $\Delta_{\perp}$ = 20.8~cm$^{-1}$ and $\Delta_{\parallel}$ = 12.7~cm$^{-1}$, with a splitting induced by anisotropy. (ii) The triplet excitations show a dispersion up to a maximum energy of $\sim 2.4 J$ ($\sim 6.7 \Delta$) at the zone boundary ($q_{\mathrm{chain}} = \pi /2$ and 3$\pi / 2$). (iii) Interchain interactions and anisotropies are given by $J_{\mathrm{inter}} \approx 0.26$~meV and $D = -0.29$~meV~\cite{bera-15b}. (iv) At the AF zone center the resulting splitted branches are at 1.56 $\pm$ 0.1~meV (12.64 $\pm$ 0.81~cm$^{-1}$) and 2.58 $\pm$ 0.1~meV (20.90 $\pm$ 0.81~cm$^{-1}$), respectively~\cite{bera-13, bera-15, bera-15b}.

%$g = 2.24$, $\Delta_{\perp} = 2.58 meV = 20.8 cm$^{-1}$

%Theoretically predicted intrinsic Haldane gap is $\Delta_{0}$ = 0.41J = 0.41∙8.7 meV = 3.57 meV (28.6~cm$^{-1}$).

%Complex interchain couplings lead to three gap energies 2.2, 3.6, 6.1 meV (17.6, 28.8, and 48.8) at different AFM zone centers and each of these modes splits into two branches of gapped excitations by anisotropy: 1.57 and 2.58 meV; 3.07 and 3.76 meV; 5.78 and 6.18 meV.

% $g = 2.24$, $\Delta_{\perp} = 20.8 \pm 0.1$ cm$^{-1}$
% $\Delta_{\parallel} = 12.7\pm 0.1$ cm$^{-1}$.
% The anisotropy induced triplet splitting is $\Delta_{\perp} - \Delta_{\parallel} = 8.1$ cm$^{-1}$.
%The ESR experiments~\cite{wang-13} show a slightly different value of $\Delta_{\perp} - \Delta_{\parallel}$ splitting of about 4.67 cm$^{-1}$. The gap $\Delta_{\perp} = 22$ cm$^{-1}$ corresponds to our experimental data.

%In the supplement, a modeling of ESR and neutron scattering data is used to reconfirm these parameters independently.

\begin{figure*}
	\label{figure2}
	\centering
	\includegraphics[width=16cm]{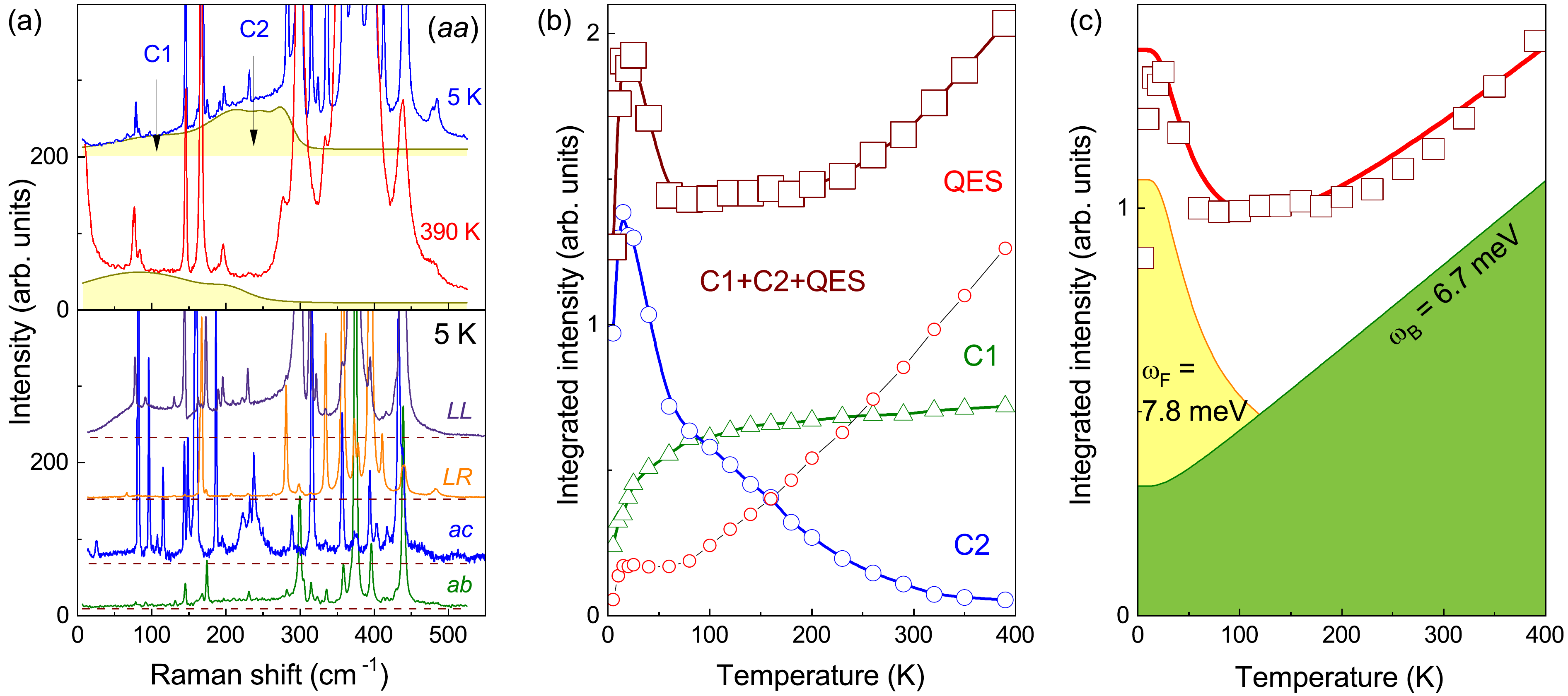}
	\caption{(Color online) (a) Raman spectra of SrNi$_2$V$_2$O$_8$ taken perpendicular to the Haldane chain direction in ($aa$) scattering geometry at 7 K and 390 K together with low-temperature spectra in ($LL$), ($LR$) and ($ab$) polarization. The yellow-shaded backgrounds represent the unconventional scattering continuum with two maxima, $C1$ and $C2$. (b) Integrated intensities of the $C1$, $C2$, quasielastic scattering (QES, E$\approx$0), and their sum. The lines are guides to the eye. (c) Decomposition of the sum from (b) (squares) into contributions with Fermionic and Bosonic statistics (full red line). Individual components are given as shaded regions. }
\end{figure*}

%%%%%%%%%%%%%%%%%%%%%%%%%%%%%%%%%%%%%%%%%%%%%
%\section{Experimental details}
%%%%%%%%%%%%%%%%%%%%%%%%%%%%%%%%%%%%%%%%%%%%%

Experimental details of the SrNi$_2$V$_2$O$_8$ single crystal growth and characterization have been reported earlier~\cite{bera-15}. Raman scattering experiments were performed in a quasi-backscattering geometry on $ab$ and $ac$ surfaces with the excitation wavelength $\lambda = 532.1$ nm and a power level $P = 5$ mW in parallel [($aa$), ($cc$)], crossed [($ab$), ($ac$)], and circular [(LR), (LL)] light polarizations. Single crystal orientation has been independently assured using the anisotropy of phonon Raman scattering and Laue diffraction method. Raman spectra were measured in a variable temperature closed cycle cryostat and a 10 T superconducting solenoid with the magnetic field along the $c$ axis of the crystal.

%%%%%%%%%%%%%%%%%%%%%%%%%%%%%%%%%%%%%%%%%%%%%
%\section{Conventional magnetic Raman scattering}
%%%%%%%%%%%%%%%%%%%%%%%%%%%%%%%%%%%%%%%%%%%%%

Polarized Raman spectra of SrNi$_2$V$_2$O$_8$ show all expected phonon modes and  their small linewidth and well-defined selection rules indicate the high sample quality. There are no significant phonon anomalies as function of temperatures, and therefore no evidence for a structural instability~\cite{Phonons}.

At low temperatures (T$<$30K) and in (ac) scattering geometry we notice two modes, at 22 cm$^{-1}$ (blue arrow) and at 220 cm$^{-1}$ (red arrow), see Fig. 1(a) and a summary in Table 1. The ($ac$) component with one polarization component along the chain direction corresponds to $E$ symmetry.  The first mode has a rather narrow linewidth and its intensity $I(T)$ is critically depressed with increasing temperatures, see Fig. 1(b), blue symbols. Also an applied transverse magnetic field leads to a characteristic shift, see Fig. 1(c). This shift corresponds very well to previous neutron scattering and ESR data~\cite{bera-13, bera-15, bera-15b, wang-13} of triplet excitations in magnetic fields. Therefore we assign this mode to a one-triplet excitation.

The second mode has a factor 6 larger linewidth. It does not show a pronounced frequency shift with temperature and $I(T)$ is only fully depleted for temperatures above $\sim 30$ K, i.e. it is less sensitive to thermal fluctuations. The mode's energy is larger than the maximum energy of the neutron triplet dispersion ($\sim 23$ meV, 185 cm$^{-1}$)~\cite{bera-15}. Nevertheless, it fits well to $3J = 210$ cm$^{-1}$.

A similar mode at $3J$ has also been observed in the Haldane chain compound Y$_2$BaNiO$_5$~\cite{sulewski-95}. The factor $3$ can be rationalized in an Ising-like spin system considering breaking spin bonds by a spin exchange process. This scheme leads to a typical energy $E_{\mathrm{max}}=(z\cdot 2S - 1)J$~\cite{hayes-78, Cottam-Lockwood}. With $S = 1$ and the coordination number, $z = 2$, $E_{\mathrm{max}}=3J$ is derived. The good agreement of such a heuristic picture of an Ising-like, N\'{e}el ground state to an excitation in a quantum disordered system is far from trivial. In $1D$ or $2D$ spin liquids a considerable renormalization to lower energies and increasing linewidth is observed~\cite{lemmens-03}. This could imply well-defined quasiparticles and weak fluctuation corrections in this Haldane system. However, it might also mean that these excitations are stabilized by some other mechanism, as discussed later.

We use $I(T) \propto (1-e^{\Delta/k_BT})$, with singlet-triplet gap $\Delta=3.6$ meV, to fit the one-triplet intensity (blue symbols), see solid curve in Fig. 1 (b). This fit corresponds to a depression of $I(T)$ due to scattering on thermally populated triplets across this gap, i.e. a bosonic statistics. Such a model works very well for SrCu$_2$(BO$_3$)$_2$ with orthogonal spin dimers and strongly localized excited triplets~\cite{lemmens00}.

The two-triplet scattering involves double exchange processes averaging over a regime of $k$ space around the zone boundary. Therefore a simple analytic expression of $I(T)$ does not exist~\cite{Cottam-Lockwood}. The observed smoother dependence is rationalized by a superposition of scattering processes with a dispersing singlet-triplet gap. A second scenario that is relevant for low-dimensional systems considers the importance of local spin-spin correlations. Therefore thermally induced triplet states do not scatter efficiently~\cite{lemmens-03,lemmens00}.

%Due to the screw-like structure of the Ni$^{2+}$ chains, the Brillouin zone (BZ) boundary triplet excitations of the Haldane chain are shifted into the center of the reduced BZ.

%%%%%%%%%%%%%%%%%%%%%%%%%%%%%%%%%%%%%%%%%%%%%
%\subsection{Unconventional Raman scattering}
%%%%%%%%%%%%%%%%%%%%%%%%%%%%%%%%%%%%%%%%%%%%%

%use a different figure that shows the continuum in different polarizations. Use origin, copy to Corel Draw and make a preliminary figure

An unconventional Raman signal is observed between 100 and 280 cm$^{-1}$ with broad maxima and a linewidth of $\approx$200 cm$^{-1}$. Its onset energy is around 60 cm$^{-1}$ and a high energy cut-off exists at $\approx 350$ cm$^{-1}$, see Fig. 2 (a). This signal is only observed in ($LL$) and ($aa$) polarization and not in ($LR$), ($ab$), and ($ac$). Such a strict selection rule points to an intrinsic origin and is not compatible with fluorescence or phonon density of states. Therefore we attribute the continuum to some sort of magnetic/electronic scattering.

% appendix with resonance Raman, etc studies

After subtracting spectral contributions due to phonons we have identified three major contributions, i.e. lower and higher energy maxima ($C1$,$C2$) and quasielastic scattering ($QES$). The respective integrated intensities and their sum are given in Fig. 2 (b). Spectral weight from the high energy $C2$ to lower energy $C1$ contribution is transferred into QES with increasing temperatures. At lowest temperatures, comparable to the Haldane gap ($\Delta~\approx$ 42~K), there is a general drop of intensity. Furthermore, a magnetic field leads to a suppression of the scattering intensity, see Appendix. For larger magnetic fields SrNi$_2$V$_2$O$_8$ shows a quantum phase transition into a N\'eel-ordered phase.

We notice that the sum of these intensities reveals no appreciable temperature dependence, at least in the temperature range 50-230~K. This is in sharp contrast to $I(T)$ of one-triplet and two-triplet scattering (Fig. 1 (b)). However, Raman scattering on quantum spin liquids based on the Heisenberg Kagome~\cite{wulferding-10} and the Kitaev honeycomb lattice~\cite{sandilands-15,glamazda-16} show a very similar weak temperature dependence. The latter is attributed to their inherent entanglement that make thermal fluctuations ineffective.

%of fractional excitations emerging from .
% what means LL polarization .. and aa with respect to exchange

Also the observation of this mode in ($aa$) polarization with a high energy cut-off is rather unexpected. This polarization has no component along the chain direction with large exchange coupling. The Loudon-Fleury approach implies that the spatial correlations of the exchange integral are probed by the selected light polarization vectors. Due to the small interchain coupling two-particle scattering probed in polarizations perpendicular to the chains should have an energy scale of only 2$\Delta_{\perp}\approx$~40 cm$^{-1}$. On the other hand SrNi$_2$V$_2$O$_8$ is rather close to a phase boundary to Ising order due to anisotropy and interchain interaction. Strong fluctuations at such a boundary could favor a more complex mechanism of light scattering beyond the basic Loudon-Fleury approach.

Therefore, we propose an alternative, Floquet-like scenario that consists of a two step process: In a first step, the incident light drives the system into an excited state where spins between neighboring chains are exchanged [e.g., along $J_{\mathrm{inter}}$, see Fig. 1(d)]. These states consists of defects within the topological string order. In the second step these states relax along the chain leading to a high energy scale, comparable to the energy of dispersing triplets. This scenario introduces ''edge" states within the topologically protected bulk due to the light-induced spin exchange. The first step can also be replaced by a thermally induced spin-flip. This could be the origin of the spectral weight shift in the light scattering process at elevated temperatures \cite{becker-17}. Such a complex exchange-induced, topological Raman process may also lead to a temperature dependence different from the expected Bose statistics.

%%%%%%%% as mentioned above the selection rules and light scattering polarization give informations about the essential correlation process that may not be the dominant one as shown for 2MRS in pure chains, dominated by nnn or dimerization

To determine the intensity $I(T)$ of the broader continuum we subtract the phonon contributions and fit the remaining spectral weight to a sum of Gaussians, see large squares in Fig. 2(c). The resulting temperature dependence with a minimum at around 50 K and an increase both towards higher and lower temperatures is completely different from conventional magnetic Raman scattering (as discussed below). Motivated by this difference and the above discussed Raman scattering process we have deconvoluted $I(T)$ into a sum of a two-Fermion and a single Bose contribution. This corresponds to an approach used for fractionalized modes in spin liquid states of Kitaev systems \cite{nasu-16}. It takes into account that spin liquids carry topological defects due to the creation/annihilation of pairs of Fermions (Majorana Fermions). The bosonic contribution corresponds to their binding by residual interactions or to additional low-energy triplet. For the Kitaev spin liquid materials $\alpha$-RuCl$_3$ and different phases of Li$_2$IrO$_3$ such a deconvolution leads to a remarkable description of $I(T)$~\cite{nasu-16,sandilands-15, knolle-13, knolle-14, perreault-15, nasu-14}.

%In these cases the two-Fermion and Boson contributions are of similar weight.

In detail, the deconvolution consists of a sum of two-Fermion $[(1-f(\omega_F)]^2$ and Bose contributions ($1 + n(\omega_B)) = 1/[1 - \exp(-\hbar\omega_B/k_BT )]$, with $\omega{_F}=7.8 \pm 1$~meV and $\omega{_B}=6.7 \pm 1$~meV, see red line in Fig. 2(c). The respective weighting factors are proportional to the height of the two shaded areas, see dashed lines. Both characteristic energies, $\omega{_F}$ and $\omega_B$, are of the order of the doubled Haldane gap of SrNi$_2$V$_2$O$_8$ ($\Delta=3.6$ meV) giving evidence for their intrinsic origin. Their order of magnitude is comparable to the ones of spin liquid materials mentioned in the previous paragraph.

%, while $\omega_B \approx 200$~cm$^{-1}$ is close to the intrinsic two-triplet excitation energy (see below).

If we want to rationalize this deconvolution it should be noted that the bulk of the Haldane phase is based on a very local product state with a gap and no long range entanglement. In contrast to defects and edges, the bulk does not support fractionalized excitations or edge modes with fermionic statistics. As we have no evidence for defects from magnetic susceptibility data we propose that the measurement process itself (exchange light scattering) and finite temperatures dynamically induce the defects and resulting novel states. This scenario is reminiscent of the concept of 
excited state quantum phase transitions\cite{ESQFT}. Both extracted energy scales, $\omega_F$ and $\omega_B$, strongly support this assumption. Furthermore, the high cut-off energy $E_{\mathrm{cut-off}}$ and the large linewidth of the order of 5-6$J$ are not compatible with the coordination number $z=2$ of a 1D chain.

Within the scheme of exchange-induced, broken bonds a maximum at about 3$J$ would be expected, in excellent agreement with the observed broad maxima. The strict selection rules of the continuum with all components perpendicular to the chain evidence that the existing correlations are rather two-dimensional contrasting the dominating chain exchange that exists in the ground state. On the other side, recent Floquet modelling of periodically driven Mott insulators show nonequilibrium steady states with an enhancement of correlations \cite{Peronaci-20}.

Finally, we note that our Raman experiments on the isostructural $S=1/2$ analogue SrCo$_2$V$_2$O$_8$ shows an in-chain, spinon continuum in good agreement with neutron scattering data \cite{Bera-17}. There is no evidence for high energy scattering with out-of chain polarizations. The corresponding data is shown in the Appendix.

\begin{table*}
	\caption{\label{tab:table1}Summary of the observed magnetic excitations including energy at the maximum of scattering intensity, polarization, linewidth ($\Gamma$). In addition we give the polarization of quasielastic scattering.}
	\begin{ruledtabular} \begin{tabular}{c|c|c|c|c|c}
		Nr& $E_{\mathrm{max}}$ (cm$^{-1}$)&$\Gamma$ (cm$^{-1}$) &Polarization&Parameters&Assignment, comments\\ \hline
		1 & 22        &2.5 			 &($ac$)		    & $\Delta=3.6$ meV (29 cm$^{-1}$) & 1-triplet, B-dep. \\
		2 & 220       &15 			 &($ac$)		    & $E_{\mathrm{max}}\approx$$3J$   & 2-triplet, no B-dep. \\
		3 & 100 - 280 &$\approx$ 200 &($\textbf{aa}$, $LL$, $ac$)&0-6$J$, $\omega{_F}$=7.8~meV & unconventional continuum \\
		4 & $\approx$0&- 			 &($aa$)  	        & -          & quasielastic \\
	\end{tabular} \end{ruledtabular} \end{table*}

We summarize the properties of all three modes attributed to magnetic Raman scattering in Table 1. They clearly differ with respect to energy, linewidth, polarization, and in their dependence on temperature, as well as on an external magnetic field.

\section{Summary}

The presented polarization-resolved Raman spectra of single crystalline SrNi$_2$V$_2$O$_8$ provide a view of the manifold of excited states in a quasi-one-dimensional $S = 1$ Haldane system close to a critical state. Besides a one-triplet and a two-triplet mode, a broad continuum is observed that we attribute to a novel exited state Raman process induced by the exchange process itself. The proposed scattering mechanism involves the generation of a light-induced topological interchain-defect that relaxes via a high energy, intrachain process. We have discussed the high cut-off energy of this process to be related to enhanced dimensionality (2D) or a Floquet-like enhancement of electronic correlations.

\begin{acknowledgments}
We acknowledge important discussions with W. Brenig, J. Knolle, and S. R. Manmana. This research was funded by the DFG Excellence Cluster QuantumFrontiers, EXC 2123, as well as by DFG Le967/16-1, DFG-RTG 1952/1, and the Quantum- and Nano-Metrology (QUANOMET) initiative within project NL-4. D.W. acknowledges support by the Institute for Basic Science 
(IBS-R009-Y3).
\end{acknowledgments}

\section{Appendix}

\begin{figure*}
\label{figure5}
\centering
\includegraphics[width=12cm]{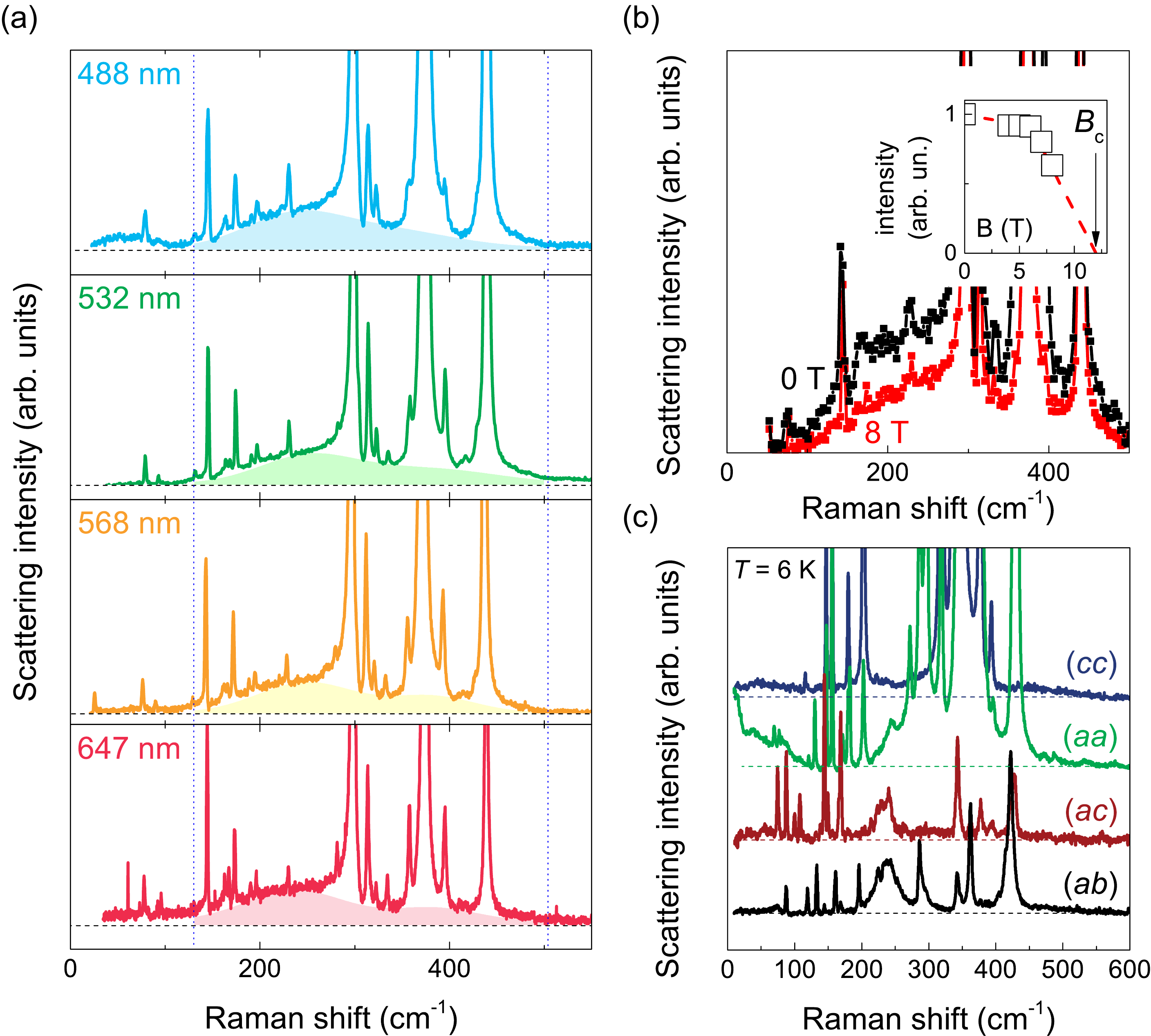}
\caption{(Color online) (a) Raman data obtained at $T=6$ K and in ($ab$) polarization using various laser energies. (b) Magnetic field ($B$) dependence of the scattering continuum in ($ab$) polarization. The inset plots its integrated intensity over $B$. (c) Polarization-resolved Raman data of the iso-structural $S=1/2$ system SrCo$_2$V$_2$O$_8$.}
\end{figure*}

In this appendix we investigate whether resonance effects or the skewed spin chains formed by edge-sharing NiO$_6$ octahedra along the $c$ axis are responsible for the broad continua of magnetic/electronic Raman scattering. Therefore we did detailed investigations using Lasers of different photon energies. Furthermore, we studied the related, Co based SrCo$_2$V$_2$O$_8$ system with similar local coordinations.

In Fig. 3 (a) four data sets are given with the indicated wavelength of the incident Laser radiation. The general shape and spectral weight continuum does not evidence any effect. This supports our assumption that it is of intrinsic origin and not related to other (e.g. fluorescence) processes. We also note that the continuum's intensity shows no significant laser energy dependence. In Fig. 3(b) the intensity of the continuum as a function of magnetic field is plotted. Here the magnetic field is aligned along the crystallographic $c$ axis, parallel to the chain direction. Since the mode is sensitive to external magnetic fields, we can assume that it is of magnetic origin. Furthermore, its intensity is strongly decreased for fields approaching 12 T, which corresponds to the critical field. At larger fields, full spin polarization is achieved~\cite{bera-15b}.

In the iso-structural sister compound SrCo$_2$V$_2$O$_8$ we find no evidence for a comparable broad continuum with a polarization perpendicular to the chain direction [see Fig. 3(c)]. SrCo$_2$V$_2$O$_8$, with comparable intrinsic energy scales, is the $S=1/2$ analogue to our title compound. Hence, its excitation spectrum is gapless and no topology-related edge states exist, in accordance with our data. Taken together, these observations strongly support our interpretation of the broad signal emerging in Raman scattering experiments on SrNi$_2$V$_2$O$_8$.

% !!!!!!!!!! Further describe the data

\end{document}